\documentclass[journal]{IEEEtran}

\usepackage{amsmath,amssymb,amscd,latexsym,dsfont,psfrag,mathrsfs,pstricks}
\usepackage{tabularx,multirow}
\usepackage{graphicx,color,array,cite}
\usepackage{float}
\usepackage{tikz}

\newcommand{\rect}{\mathop{\mathrm{rect}}\nolimits}
\makeatletter

\newcommand{\Rmnum}[1]{\expandafter\@slowromancap\romannumeral #1@}
\makeatother
\newcommand{\aaa}{0.81649658} 
\newcommand{\bbb}{0.57735027} 
\newcommand{\ccc}{0.8660254}  

\newcommand{\outt}{0}
\newcommand{\inn}{0}

\newcommand{\hsp}{12pt}
\begin{document}
\title{A Two-Dimensional Signal Space for Intensity-Modulated Channels}
\author{Johnny~Karout,~\IEEEmembership{Student Member,~IEEE,}
        Gerhard~Kramer,~\IEEEmembership{Fellow,~IEEE,}
                Frank~R.~Kschischang,~\IEEEmembership{Fellow,~IEEE,}
        and~Erik~Agrell
        \thanks{J.~Karout and E.~Agrell are with the Department of Signals and Systems, Chalmers University of Technology, SE-412 96 Gothenburg, Sweden ({johnny.karout, agrell}@chalmers.se).}
        \thanks{G.~Kramer is with the Institute for Communications Engineering, Technische Universit\"at M\"unchen,
D-80290 Munich, Germany (gerhard.kramer@tum.de).}
        \thanks{F.~R.~Kschischang is with the Edward S. Rogers Sr. Department of Electrical and Computer Engineering, University of Toronto, Toronto, ON M5S 3G4, Canada (e-mail: frank@comm.utoronto.ca). His work was performed while on a Fellowship at the Institute for Advanced Study, Technische Universit\"at M\"unchen, Lichtenbergstrasse 2a, D-85748 Garching, Germany.}
}    
\maketitle

\begin{abstract}
A two-dimensional signal space for intensity-modulated channels is presented. Modulation formats using this signal space are designed to maximize the minimum distance between signal points while satisfying average and peak power constraints. The uncoded, high-signal-to-noise ratio, power and spectral efficiencies are compared to those of the best known formats. The new formats are simpler than existing subcarrier formats, and are superior if the bandwidth is measured as 90\% in-band power. Existing subcarrier formats are better if the bandwidth is measured as 99\% in-band power.
\end{abstract}


\IEEEpeerreviewmaketitle


\section{Introduction}

\IEEEPARstart{I}{ntensity} modulation with direct detection (IM/DD) is widespread for low-cost optical communication systems, e.g., wireless optical links~\cite{Barry1994,Kahn1997,Hranilovic2004a} and short-haul fiber links~\cite{Randel2008}. 
IM/DD permits only the intensity of light to carry information. In contrast, coherent optical systems such as long-haul fiber links let data modulate the optical carrier's amplitude and phase via, e.g., $M$-ary quadrature amplitude modulation ($M$-QAM).
Designing IM/DD formats with good power and spectral characteristics is challenging~\cite{Barry1994,Kahn1997, Hranilovic2003,Karout2011GC,Karout2011IT}.

In the absence of optical amplification, IM/DD systems can be modeled as additive white Gaussian noise (AWGN) channels with nonnegative inputs~\cite[Ch. 5]{Barry1994},~\cite{Kahn1997,Hranilovic2003},~\cite[Sec. 11.2.3]{Hobook2005}. Nonnegative $M$-ary pulse amplitude modulation ($M$-PAM) such as on-off keying (OOK)~\cite[Eq.~(5.8)]{Barry1994} is a natural modulation format but it is power inefficient for $M>2$~\cite{Walklin}.
Subcarrier modulation (SCM) allows using $M$-QAM by adding a direct current (DC) bias to the electrical signal to make it nonnegative~\cite[Ch.~5]{Barry1994}. The DC bias does not carry information if it is independent of the transmitted information. A signal space for IM/DD channels was presented in~\cite{Hranilovic2003} and power-efficient subcarrier modulation formats were designed. 
In our prior work, a three-dimensional signal space for IM/DD, whose signal sets are denoted as raised-QAM~\cite{Hranilovic2003}, was used to numerically optimize modulation formats for different power constraints~\cite{Karout2011GC,Karout2011IT}. 

In this work, we present a two-dimensional signal space for optical IM/DD systems. The resulting modulation formats have simpler modulator and demodulator structures than the three-dimensional formats studied before. 
Their power and spectral efficiencies are evaluated and compared to the previously best known formats. 

\section{System Model}
\begin{figure}
\begin{minipage}{1\columnwidth}
\centering
\includegraphics[width=1\columnwidth]{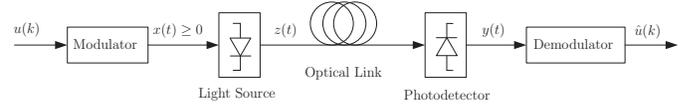}
\end{minipage}
  \caption{Passband transceiver of IM/DD systems.}
\label{fig:model}
\end{figure}

\begin{figure*}
\begin{center}

\begin{tikzpicture}[>=stealth,scale=0.75]
\coordinate (O) at (0,0);
\coordinate (A) at (\bbb,\aaa);
\coordinate (B) at (-\bbb,\aaa);
\foreach \point in {O}
   \shadedraw[shading=radial,inner color=black!\outt,outer color=black!\inn,draw=black] (\point) circle (0.5);
   
      \foreach \point in {A,B}
   \shadedraw[shading=radial,inner color=black!10,outer color=black!5,draw=black] (\point) circle (0.5);

\draw[->,thick] (-2,0) -- (2,0) node[anchor=north east] {$\phi_2(t)$};
\draw[->,thick] (0,-0.75) node[anchor=north] {\raisebox{0ex}[2ex][0ex]{$\mathscr{T}_{\bar{P}_o,3}$}}
 -- (0,3.5) node[anchor=west] {$\phi_1(t)$};
 
\draw[black,very thick] (2,2.8284271) -- (0,0) -- (-2,2.8284271);
\foreach \point in {O,A,B}
   \filldraw[fill=black,draw=black!50] (\point) circle (0.06);
\end{tikzpicture}
\hspace{\hsp}
\begin{tikzpicture}[>=stealth,scale=0.75]
\coordinate (O) at (0,0);
\coordinate (A) at (0.5,\ccc);
\coordinate (B) at (-0.5,\ccc);
\foreach \point in {O}
   \shadedraw[shading=radial,inner color=black!\outt,outer color=black!\inn,draw=black] (\point) circle (0.5);
   
   \foreach \point in {A,B}
   \shadedraw[shading=radial,inner color=black!10,outer color=black!5,draw=black] (\point) circle (0.5);

\draw[->,thick] (-2,0) -- (2,0) node[anchor=north east] {$\phi_2(t)$};
\draw[->,thick] (0,-0.75) node[anchor=north] {\raisebox{0ex}[2ex][0ex]{$\mathscr{T}_{\hat{P}_o,3}$}}
 -- (0,3.5) node[anchor=west] {$\phi_1(t)$};
\draw[black,very thick] (2,2.8284271) -- (0,0) -- (-2,2.8284271);
\foreach \point in {O,A,B}
   \filldraw[fill=black,draw=black!50] (\point) circle (0.06);
\end{tikzpicture}
\hspace{\hsp}
\begin{tikzpicture}[>=stealth,scale=0.75]
\coordinate (O) at (0,0);
\coordinate (A) at (\bbb,\aaa);
\coordinate (B) at (-\bbb,\aaa);
\coordinate (C) at (0,2*\aaa);
\foreach \point in {O,A,B,C}
   \shadedraw[shading=radial,inner color=black!\outt,outer color=black!\inn,draw=black] (\point) circle (0.5);
\draw[->,thick] (-2,0) -- (2,0) node[anchor=north east] {$\phi_2(t)$};
\draw[->,thick] (0,-0.75) node[anchor=north] {\raisebox{0ex}[2ex][0ex]{$\mathscr{T}_{4}$}}
 -- (0,3.5) node[anchor=west] {$\phi_1(t)$};
\draw[black,very thick] (2,2.8284271) -- (0,0) -- (-2,2.8284271);
\foreach \point in {O,A,B,C}
   \filldraw[fill=black,draw=black!50] (\point) circle (0.06);
\end{tikzpicture}
\hspace{\hsp}
\begin{tikzpicture}[>=stealth,scale=0.75]
\coordinate (O) at (0,0);
\coordinate (A) at (\bbb,\aaa);
\coordinate (B) at (-\bbb,\aaa);
\coordinate (C) at (0,2*\aaa);
\coordinate (D) at (2*\bbb,2*\aaa);
\coordinate (E) at (-2*\bbb,2*\aaa);
\coordinate (F) at (\bbb,3*\aaa);
\coordinate (G) at (-\bbb,3*\aaa);
\foreach \point in {O,A,B,C}
   \shadedraw[shading=radial,inner color=black!\outt,outer color=black!\inn,draw=black] (\point) circle (0.5);
   
\foreach \point in {D,E}
   \shadedraw[shading=radial,inner color=black!25,outer color=black!20,draw=black] (\point) circle (0.5);
   
\foreach \point in {F,G}
   \shadedraw[shading=radial,inner color=black!50,outer color=black!40,draw=black] (\point) circle (0.5);

\draw[->,thick] (-2,0) -- (2,0) node[anchor=north east] {$\phi_2(t)$};
\draw[->,thick] (0,-0.75) node[anchor=north] {\raisebox{0ex}[2ex][0ex]{$\mathscr{T}_{\bar{P}_o,8}$}}
 -- (0,3.5) node[anchor=west] {$\phi_1(t)$};
\draw[black,very thick] (2,2.8284271) -- (0,0) -- (-2,2.8284271);
\foreach \point in {O,A,B,C,D,E,F,G}
   \filldraw[fill=black,draw=black!50] (\point) circle (0.06);
\end{tikzpicture}
\hspace{\hsp}
\begin{tikzpicture}[>=stealth,scale=0.75]
\coordinate (O) at (0,0);
\coordinate (A) at (\bbb,\aaa);
\coordinate (B) at (-\bbb,\aaa);
\coordinate (C) at (0,2*\aaa);
\coordinate (D) at (1.0774,1.6825);
\coordinate (E) at (-1.0774,1.6825);
\coordinate (F) at (0.5,2.4990);
\coordinate (G) at (-0.5,2.4990);
\foreach \point in {O,A,B,C}
   \shadedraw[shading=radial,inner color=black!\outt,outer color=black!\inn,draw=black] (\point) circle (0.5);

\foreach \point in {D,E}
   \shadedraw[shading=radial,inner color=black!25,outer color=black!20,draw=black] (\point) circle (0.5);
   
\foreach \point in {F,G}
   \shadedraw[shading=radial,inner color=black!50,outer color=black!40,draw=black] (\point) circle (0.5);
\draw[->,thick] (-2,0) -- (2,0) node[anchor=north east] {$\phi_2(t)$};
\draw[->,thick] (0,-0.75) node[anchor=north] {\raisebox{0ex}[2ex][0ex]{$\mathscr{T}_{\hat{P}_o,8}$}}
 -- (0,3.5) node[anchor=west] {$\phi_1(t)$};
\draw[black,very thick] (2,2.8284271) -- (0,0) -- (-2,2.8284271);
\foreach \point in {O,A,B,C,D,E,F,G}
   \filldraw[fill=black,draw=black!50] (\point) circle (0.06);
   \end{tikzpicture}
   
\caption{Two-dimensional constellations optimized for average and peak optical power. 
Circles which are different between constellations of the same size are shaded differently.}
\label{fig:const_all}
\end{center}
\vspace{-0.6cm}
\end{figure*}
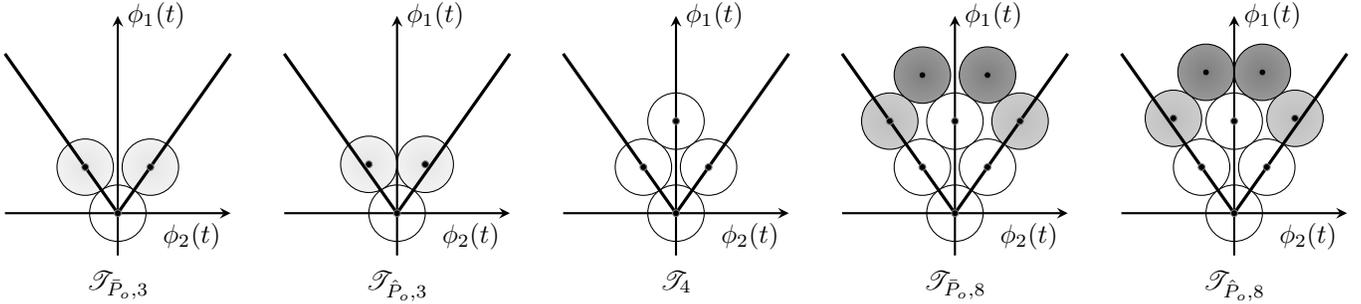

Consider an AWGN channel whose input $x(t)$ is nonnegative. 
The symbols $u(k)$, for $k=\ldots,-1,0,1,\ldots$, are independent and uniformly distributed over $\{0,1,\ldots,M-1\}$.
The modulator maps each $u(k)$ to a real and nonnegative waveform belonging to the signal set $S=\{s_0(t),s_1(t), \ldots, s_{M-1}(t)\}$, where $s_i(t)=0$ for $t \notin [0,T)$, $i=0,1,\ldots,M-1$, and $T$ is the symbol period. 
The transmitted waveform is
\begin{equation}\label{sig:xt}
x(t)=\sum_{k=-\infty}^{\infty} s_{u(k)}(t-k T).
\end{equation}
The received signal is modeled as
\begin{equation}\label{eq:basebandreceivedsignal}
    y(t)= x(t) + n(t),
\end{equation}
where $n(t)$ is a zero-mean white Gaussian process with double-sided power spectral density $N_0/2$. 
The demodulator is a correlator or matched filter receiver with a minimum-distance detector, i.e., it minimizes the symbol error rate at a given signal-to-noise (SNR) ratio~\cite[Sec. 4.1]{Simon1995} and puts out $\hat u(k)$ as the estimate of $u(k)$.

The AWGN model is reasonable for IM/DD systems when the dominating noise is from the receiver itself, and not from optical amplifiers~\cite[Ch. 5]{Barry1994},~\cite{Kahn1997,Hranilovic2003},~\cite[Sec. 11.2.3]{Hobook2005}. 
A passband model for IM/DD systems is depicted in Fig.~\ref{fig:model}. The electrical nonnegative waveform $x(t)$ modulates a light source such as a laser diode. The information is carried by the intensity of the passband signal
\begin{equation}
  z(t)=\sqrt{2 c x(t)} \cos( 2 \pi f_o t +\theta(t)),
\end{equation}
where $c$ represents the electro-optical conversion factor in watts per ampere (W/A)~\cite{Cox2002}, $f_o$ is the optical carrier frequency, and $\theta(t)$ is a random phase, uniformly distributed in $[0,2\pi)$ and varying slowly with $t$ relative to the symbol rate. 
The optical signal propagates through the optical medium depicted as an optical fiber in Fig.~\ref{fig:model}, but which could also be a free-space optical link. The photodetector at the receiver, with responsivity $r$ in A/W, detects the intensity of $z(t)$. Without loss of generality, $c$ and $r$ are normalized so that the received electrical signal $y(t)$ can be written as~(\ref{eq:basebandreceivedsignal})~\cite[p. 155]{Agrawal2005}.

\section{Signal Space Analysis} \label{sec:signal_space}
The signals in $S$ can be represented as
\begin{equation}\label{eachsignal}
    s_i(t)= \sum_{k=1}^{N} s_{i,k} \, \phi_k(t)
\end{equation}
for $i=0, \ldots, M-1$, where $\{\phi_k(t)\}_{k=1}^{N}$, $N \leq M$, is a set of orthonormal basis functions~\cite{Hranilovic2003}. 
The vector representation of $s_i(t)$ with respect to these basis functions is $\mathbf{s}_i=(s_{i,1},s_{i,2}, \ldots, s_{i,N})$. We may thus alternatively represent the signal set as $S=\{\mathbf{s}_0,\mathbf{s}_1, \ldots, \mathbf{s}_{M-1}\}$.

\begin{figure*}
\begin{minipage}{0.5\textwidth}
\centering
        \includegraphics[width=1\columnwidth]{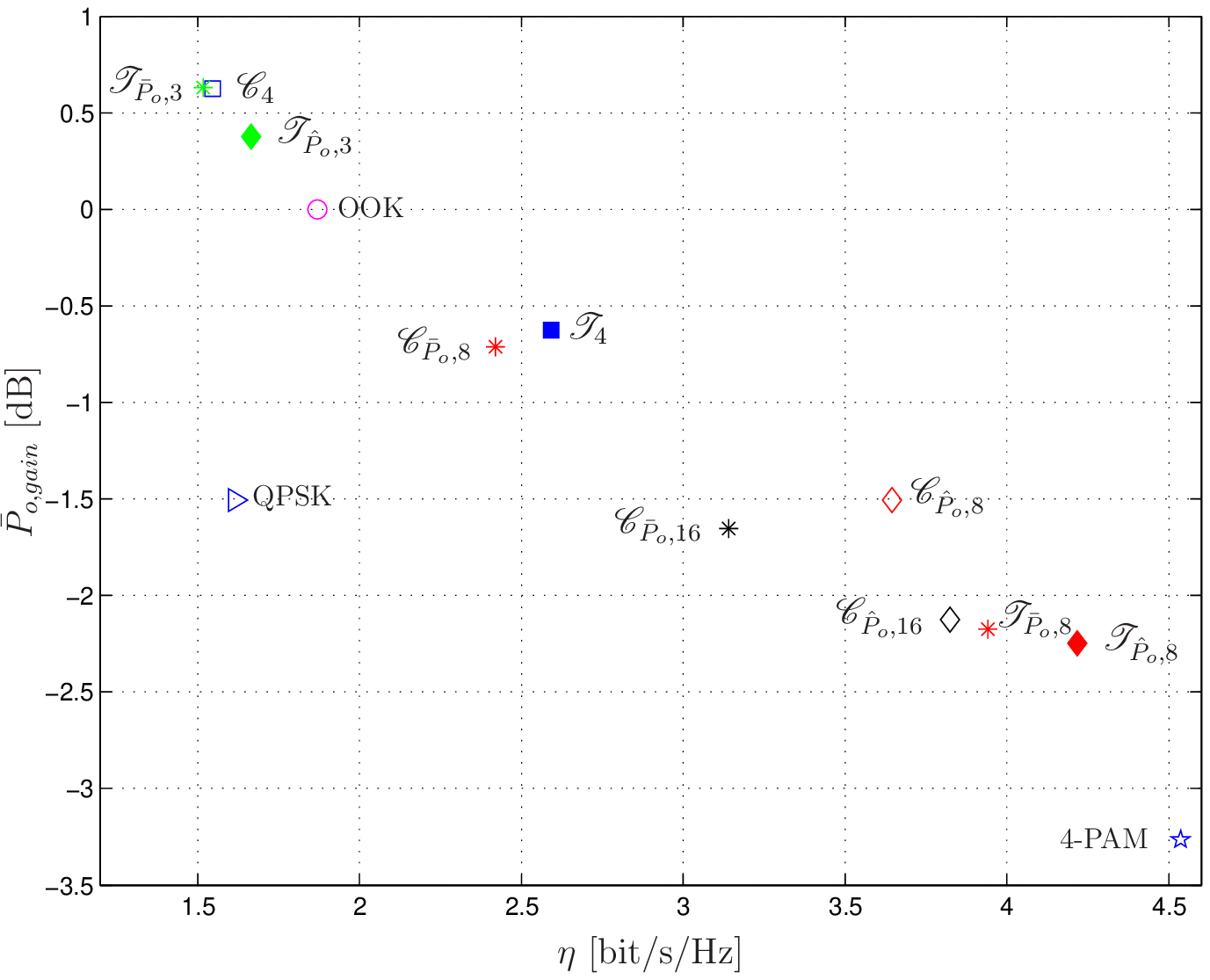}
    \caption{Asymptotic average optical power gain of modulation formats over OOK ($K=0.9$).}
    \label{fig:90:ave}
\end{minipage}%
\hspace{0.3cm}
\begin{minipage}{0.5\textwidth}
\centering
    \includegraphics[width=1\columnwidth]{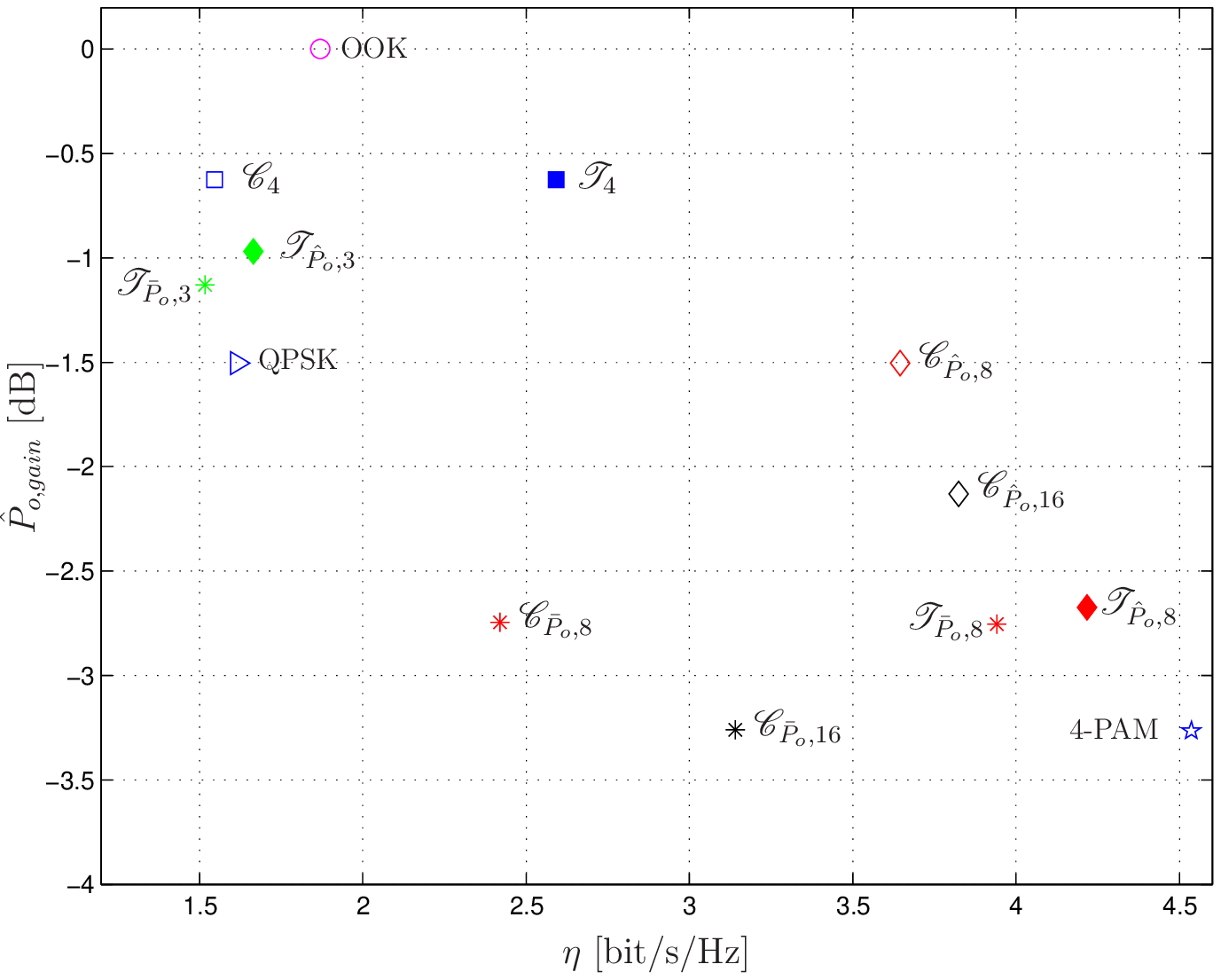}
    \caption{Asymptotic peak optical power gain of modulation formats over OOK ($K=0.9$).}
    \label{fig:90:peak}
\end{minipage}

\vspace{0.2cm}

\begin{minipage}{0.5\textwidth}
\centering
        \includegraphics[width=1\columnwidth]{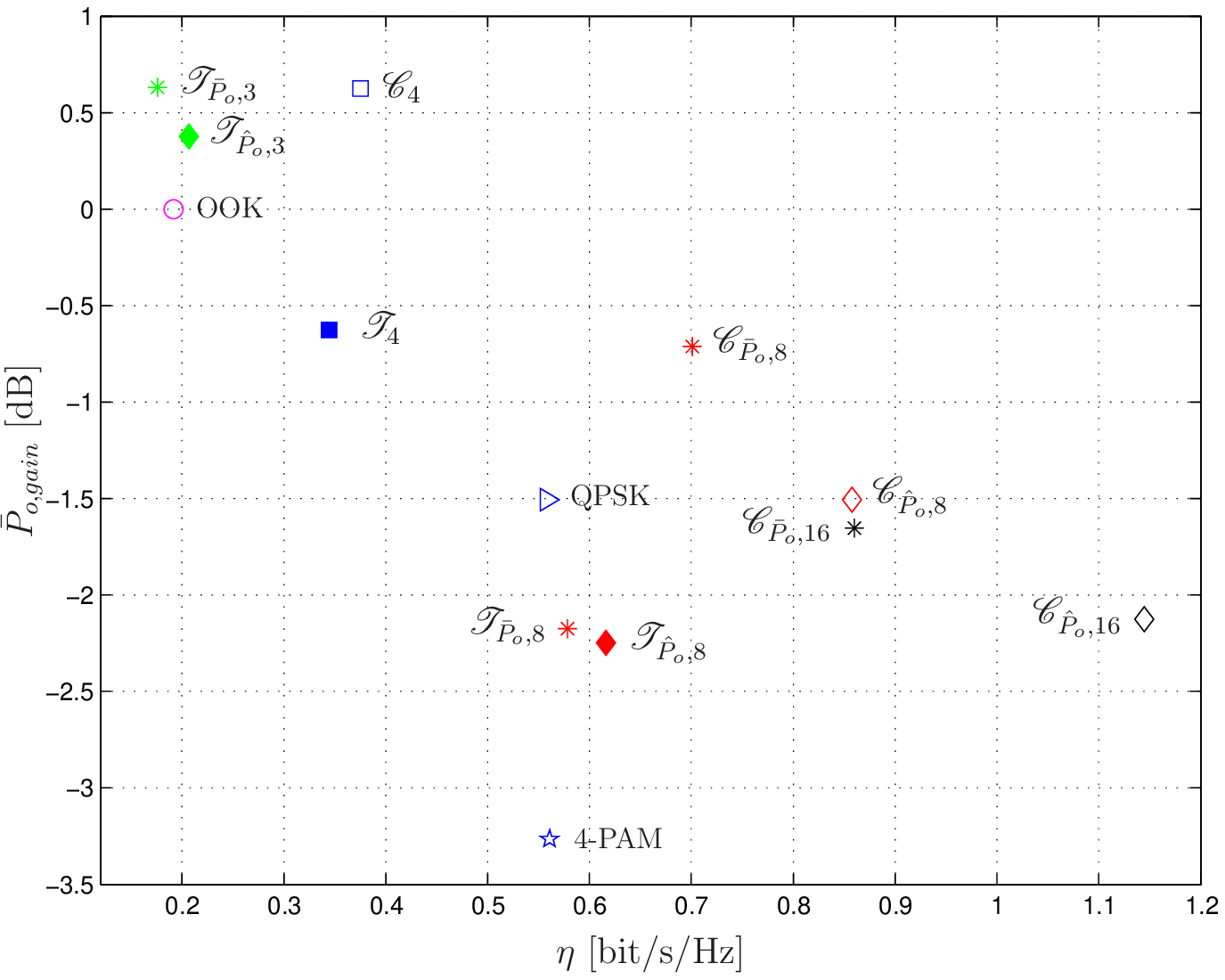}
    \caption{Asymptotic average optical power gain of modulation formats over OOK ($K=0.99$).}
    \label{fig:99:ave}
\end{minipage}%
\hspace{0.3cm}
\begin{minipage}{0.5\textwidth}
\centering
        \includegraphics[width=1\columnwidth]{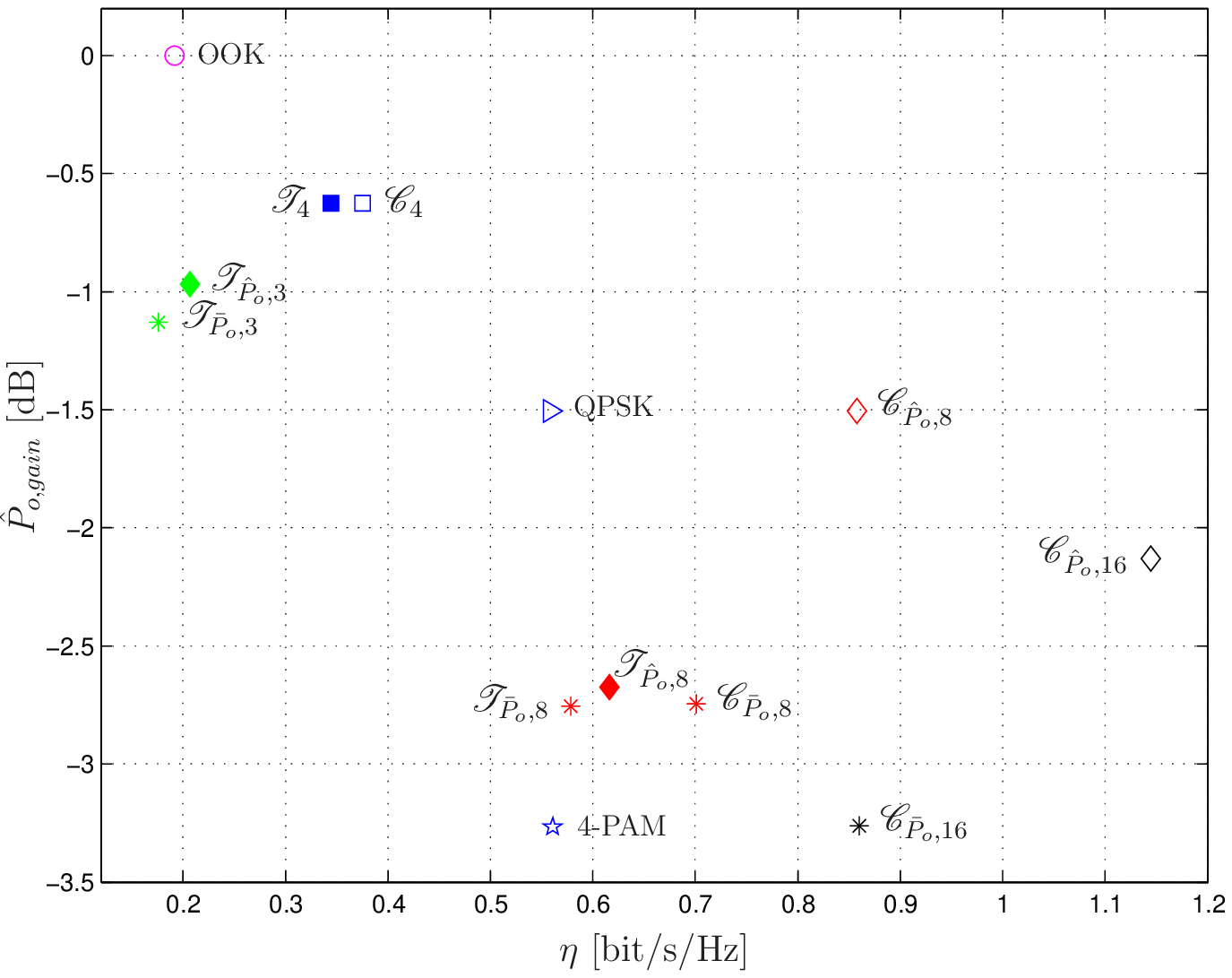}
    \caption{Asymptotic peak optical power gain of modulation formats over OOK ($K=0.99$).}
    \label{fig:99:peak}
\end{minipage}%
\vspace{-0.4cm}
\end{figure*}

Consider a two-dimensional signal space for IM/DD spanned by the basis functions
\begin{eqnarray}
 \label{eq:phi1} \phi_1(t) &=& \sqrt{ \frac{1}{T} } ~\rect\left(\frac{t}{T}\right), \\
   \label{eq:phi2} \phi_2(t) &=& \sqrt{ \frac{2}{T} }~\cos{(2\pi f t)} ~\rect\left(\frac{t}{T}\right),
\end{eqnarray}
where
\begin{eqnarray*}
    \rect(t) =
\begin{cases}
 1, & \mbox{if } 0 \leq t < 1, \\
 0, & \mbox{otherwise,}
\end{cases}
\end{eqnarray*}
and the electrical subcarrier frequency $f= 1/2T$. 
In~\cite{Hranilovic2003} and our prior work~\cite{Karout2011GC,Karout2011IT}, a three-dimensional signal space with signal sets called raised-QAM was used to design modulation formats. This three-dimensional signal space is spanned by $\phi_1(t)$ in~(\ref{eq:phi1}), and the in-phase and quadrature phase modulation formats' basis functions with $f=1/T$~\cite[Eqs.~(13)--(14)]{Karout2011IT}.
The basis function $\phi_1(t)$ represents the DC bias, and is used as in~\cite{Hranilovic2003} to guarantee signal nonnegativity.

We follow the same steps as in Theorem 1 in~\cite{Karout2011IT} for raised-QAM. The admissible region $\Upsilon$, defined as the set of two-dimensional signal vectors satisfying a nonnegativity constraint, is a two-dimensional cone with vertex at the origin, an apex angle of $\cos^{-1}(1/3)=70.528^{\circ}$, and an opening in the dimension spanned by $\phi_1(t)$.

\subsection{Example Modulation Formats}\label{sec:example_modulation}
Fig.~\ref{fig:const_all} presents several modulation formats designed for the admissible region $\Upsilon$. 
For a certain power constraint, the formats are optimized to maximize the minimum distance between points. 
As in~\cite{Hranilovic2003,Karout2011IT}, the average optical power 
 \begin{equation}\label{opticalpower}
    \bar P_{o}=  \lim_{T\to \infty}\frac{1}{2T} \int_{-T}^{T} \! z^2(t) \, \mathrm{d}t= \lim_{T \to \infty} \frac{c}{2T} \int_{-T}^{T} \!  x(t) \,  \mathrm{d}t,
  \end{equation}
and peak optical power 
\begin{equation}\label{peakoptpowersig}
  \hat{P}_o = \max_{t} \frac{z^2(t)}{2} 
   =c \max_{t} x(t)
\end{equation}
are used as design criteria. The modulation formats numerically optimized for average optical power are denoted as $\mathscr{T}_{\bar P_o,M}$, and for peak optical power as $\mathscr{T}_{\hat P_o,M}$. They are listed in Appendix~\ref{App:Const}.
For illustration, Fig.~\ref{fig:const_all} depicts the constellation points, regarded as the centers of unit-diameter circles, together with the admissible region $\Upsilon$.
$\mathscr{T}_{4}$ is a 4-ary constellation optimized for both power measures. Together with $\mathscr{T}_{\bar P_o,3}$ and $\mathscr{T}_{\bar P_o,8}$, they are subsets of a lattice where the angle between its two basis vectors is $\cos^{-1}(1/3)$, which is the apex angle of the cone.
\subsection{Performance Measures}
To evaluate performance, the uncoded and asymptotic (high-SNR) power gains with respect to OOK are considered~\cite{Kahn1997,Barry1994,Karout2011IT}. The average optical power gain $\bar P_{o,gain}$ over OOK for the same error rate performance is defined in~\cite[Eq.~(32)]{Karout2011IT}, and the peak optical power gain $\hat P_{o,gain}$ with respect to OOK is defined in~\cite[Eq.~(33)]{Karout2011IT}.
The average optical power is an important figure of merit for skin- and eye-safety in wireless optical links~\cite[Ch. 5]{Barry1994},~\cite{Kahn1997,Hranilovic2003}, and the peak power measures tolerance against nonlinearities~\cite{Inan2009}. 

The spectral efficiency measures the bit rate achieved in a given bandwidth. It is defined as
\begin{equation}
\eta=\frac{R_b}{W}
 ~\text{[bit/s/Hz]},
\end{equation}
where $R_b=R_s \log_2 M$, $R_s=1/T$ is the symbol rate in symbols per second, and $W$ is the baseband bandwidth of $x(t)$.
In~\cite{Barry1994,Karout2011GC,Karout2011IT}, $W$ was defined as the first null in the spectrum, i.e., the width of the main lobe, since most of the energy of a signal is contained in this main lobe. However, some modulation formats designed using the signal space in Sec.~\ref{sec:signal_space} have no spectral nulls. Fig.~\ref{fig:const_all} shows five formats that lack spectral nulls. This makes this definition of bandwidth misleading. 
Instead, as in~\cite[p.~49]{Hranilovic2004a}, we will use the fractional power bandwidth $W$ defined as the length of the smallest frequency interval carrying a certain fraction of the total power. Note, however, that~\cite[Eq.~(3.15)]{Hranilovic2004a} accounts only for the continuous spectrum of $x(t)$. We will include both the discrete spectrum and the continuous spectrum. For IM/DD channels, the discrete spectral component at $f=0$ (at DC) represents the average optical power of a constellation~\cite[p.~47]{Hranilovic2004a}. The fractional power bandwidth $W$ is the solution to
\begin{equation}
\frac{ \int_{-W}^{W} \! S_x(f) \, \mathrm{d} f }{ \int_{-\infty}^{\infty} \! S_x(f) \, \mathrm{d} f } =K,
\end{equation}
where $S_x(f)$ is the power spectral density of $x(t)$, and $K \in (0,1)$. 
For constellations with uniform probability distribution, $S_x(f)$ can be obtained using~\cite[Eq. (3.7.6)]{Wilson96}, which depends only on the Fourier transform of the signals in $S$.

\section{Performance Analysis} \label{sec:perf_analysis}
Figs.~\ref{fig:90:ave}--\ref{fig:99:peak} depict the (uncoded, high-SNR) average and peak optical power gains of our modulation formats with respect to OOK, and as a function of spectral efficiency $\eta$. The fractional power bandwidth $W$ was computed using $K=0.9$ and $K=0.99$. These choices of $K$ are somewhat arbitrary but are commonly used. 

In addition to the modulation formats introduced in this paper, we consider subcarrier QPSK, nonnegative 4-PAM, and three-dimensional modulation formats from previous work optimized for average optical power ($\mathscr{C}_{\bar P_o,M}$) and peak optical power ($\mathscr{C}_{\hat P_o,M}$)~\cite{Karout2011IT}. We next discuss the plots in Figs.~\ref{fig:90:ave}--\ref{fig:99:peak}. 

\subsubsection{$\bar P_{o,gain}$ vs. $\eta$ with $K=0.9$ (Fig.~\ref{fig:90:ave})}
For a fixed $M$, modulation formats optimized for average power have a larger $\bar P_{o,gain}$ than those optimized for peak power.
The three-dimensional constellation optimized for both power measures, $\mathscr{C}_{4}$, has a similar
$\eta$ and $\bar P_{o,gain}$ as $\mathscr{T}_{\bar P_o,3}$, whereas $\mathscr{T}_{4}$ has better power and spectral efficiency than $\mathscr{C}_{\bar P_o,8}$.
The three-dimensional constellations optimized for peak power, $\mathscr{C}_{\hat P_o,8}$ and $\mathscr{C}_{\hat P_o,16}$, have higher $\eta$ than $\mathscr{C}_{\bar P_o,8}$ and $\mathscr{C}_{\bar P_o,16}$. This comes at the price of a lower power efficiency.
$\mathscr{T}_{\hat P_o,8}$ has a higher $\eta$ than $\mathscr{T}_{\bar P_o,8}$, and 4-PAM has the highest $\eta$ and lowest $\bar P_{o,gain}$. 
For $K=0.9$, the modulation formats designed using the two-dimensional signal space have better spectral-efficiency characteristics than formats designed using raised-QAM. For example, $\mathscr{T}_{\hat P_o,8}$ has a better $\eta$ than $\mathscr{C}_{\hat P_o,16}$. 

\subsubsection{$\hat P_{o,gain}$ vs. $\eta$ with $K=0.9$ (Fig.~\ref{fig:90:peak})}
For a fixed $M$, modulation formats optimized for peak power have better spectral as well as power efficiency than those optimized for average power.
OOK has the best $\hat P_{o,gain}$ among all studied modulation formats. Together with $\mathscr{T}_{4}$, they have a better $\eta$ than $\mathscr{C}_{4}$, $\mathscr{T}_{\hat P_o,3}$, $\mathscr{T}_{\bar P_o,3}$, and QPSK.
The two-dimensional 8-ary formats have higher $\eta$ than the three-dimensional 16-ary formats, and 4-PAM has the highest $\eta$ and the lowest $\hat P_{o,gain}$ among the studied modulation formats. 

\subsubsection{$\bar P_{o,gain}$ vs. $\eta$ with $K=0.99$ (Fig.~\ref{fig:99:ave})}
As before, modulation formats optimized for average power have a higher $\bar P_{o,gain}$ than those optimized for peak power.
$\mathscr{T}_{\bar P_o,3}$ and $\mathscr{C}_{4}$ have the highest $\bar P_{o,gain}$ among the studied formats. QPSK and 4-PAM have similar $\eta$ which is the highest among the 2-, 3-, and 4-ary constellations.
Unlike the case where $K=0.9$, the three-dimensional 8-ary constellations are better in spectral and power efficiency in comparison to the two-dimensional 8-ary constellations. 
In addition, $\mathscr{C}_{\hat P_o,16}$ has the highest $\eta$ among the studied constellations. 
The two-dimensional constellations have more power close to DC, whereas the three-dimensional constellations have a wider main lobe.
This makes the 99\% in-band power for the three-dimensional constellations occur at frequencies lower than those for the two-dimensional ones.

\subsubsection{$\hat P_{o,gain}$ vs. $\eta$ with $K=0.99$ (Fig.~\ref{fig:99:peak})}
As in the case where $K=0.9$, OOK has the highest $\hat P_{o,gain}$.
$\mathscr{C}_{4}$ has the same power efficiency as $\mathscr{T}_{4}$, but has a higher $\eta$.
QPSK and 4-PAM, as in the previous case, have similar $\eta$ which is the highest among the 2\nobreakdash-, 3\nobreakdash-, and 4-ary constellations. The three-dimensional 8- and 16-ary constellations have higher $\eta$ than the two-dimensional formats, and the 16-ary constellation optimized for peak power has the highest $\eta$.
\section{Conclusions}
We presented a two-dimensional signal space for IM/DD that provides a good trade-off between implementation complexity and spectral efficiency. 
This signal space suggests simpler modulator and demodulator structures than for the three-dimensional raised-QAM signal space.
For a fractional power bandwidth of $K=0.9$, the two-dimensional formats have better spectral characteristics than the three-dimensional ones. Therefore, the two-dimensional formats are a good choice for single-wavelength optical systems.
However, for a fractional power bandwidth of $K=0.99$, the three-dimensional formats have better spectral characteristics. Therefore, they are suitable for wavelength-division multiplexing (WDM) systems where crosstalk between adjacent channels is important. 
For both signal spaces, modulation formats optimized for peak power achieve a higher spectral efficiency than those optimized for average power.
\appendices
\section{Obtained Constellations}\label{App:Const}
The coordinates of the optimized constellations in Sec.~\ref{sec:example_modulation} are listed below. The constellations are normalized to have unit minimum distance.
 \begin{align*}
 \mathscr{T}_{\bar P_o,3}&=\{
 (0      ,   0), (\sqrt{2/3},         1/\sqrt{3}),(\sqrt{2/3} ,   -1/\sqrt{3}) \}. \\
%
\mathscr{T}_{\hat P_o,3}  &=\{
(0, 0),( \sqrt{3}/2,1/2),(  \sqrt{3}/2, -1/2)\}. \\    
\mathscr{T}_{4}&= \mathscr{T}_{\bar P_o,3} \cup
\{  (2\sqrt{2/3}  , 0)\}. \\
\mathscr{T}_{\bar P_o,8} &=\mathscr{T}_{4}\cup  \{
       (2\sqrt{2/3}, 2/\sqrt{3}),
        (2\sqrt{2/3}, -2/\sqrt{3}),
\\      &\qquad  (\sqrt{6},         1/\sqrt{3}),
        (\sqrt{6},         -1/\sqrt{3}) \}. \\
\mathscr{T}_{\hat P_o,8} &=\mathscr{T}_{4} \cup \{
(\sqrt{2/3} + \sqrt{3}/2 ,1/2 + 1/\sqrt{3}),   \\&\qquad
(\sqrt{2/3} + \sqrt{3}/2 ,-1/2 - 1/\sqrt{3}),
\\&\qquad (2 \sqrt{2/3} + \sqrt{3}/2,1/2),
(2 \sqrt{2/3} + \sqrt{3}/2,-1/2)\}.
\end{align*} 
\section*{Acknowledgments}
J.~Karout and E.~Agrell were supported by SSF under grant RE07-0026. G.~Kramer was supported by an Alexander von Humboldt Professorship endowed by the German Federal Ministry of Education and Research. F.~R.~Kschischang was supported by a Hans Fischer Senior Fellowship of the Institute for Advanced Study, Technische Universit\"at M\"unchen, funded by the German Excellence Initiative. The authors would like to acknowledge R.~Krishnan for comments about the paper.
\bibliographystyle{IEEEtran}
\bibliography{RefIEEEJ}

\begin{thebibliography}{10}
\providecommand{\url}[1]{#1}
\csname url@samestyle\endcsname
\providecommand{\newblock}{\relax}
\providecommand{\bibinfo}[2]{#2}
\providecommand{\BIBentrySTDinterwordspacing}{\spaceskip=0pt\relax}
\providecommand{\BIBentryALTinterwordstretchfactor}{4}
\providecommand{\BIBentryALTinterwordspacing}{\spaceskip=\fontdimen2\font plus
\BIBentryALTinterwordstretchfactor\fontdimen3\font minus
  \fontdimen4\font\relax}
\providecommand{\BIBforeignlanguage}[2]{{%
\expandafter\ifx\csname l@#1\endcsname\relax
\typeout{** WARNING: IEEEtran.bst: No hyphenation pattern has been}%
\typeout{** loaded for the language `#1'. Using the pattern for}%
\typeout{** the default language instead.}%
\else
\language=\csname l@#1\endcsname
\fi
#2}}
\providecommand{\BIBdecl}{\relax}
\BIBdecl

\bibitem{Barry1994}
J.~R. Barry, \emph{{Wireless Infrared Communications}}.\hskip 1em plus 0.5em
  minus 0.4em\relax Norwell, MA, USA: Kluwer Academic Publishers, 1994.

\bibitem{Kahn1997}
J.~M. Kahn and J.~R. Barry, ``{Wireless infrared communications},''
  \emph{Proceedings of the IEEE}, vol.~85, no.~2, pp. 265--298, 1997.

\bibitem{Hranilovic2004a}
S.~Hranilovic, \emph{{Wireless Optical Communication Systems}}.\hskip 1em plus
  0.5em minus 0.4em\relax New York: Springer, 2005.

\bibitem{Randel2008}
S.~Randel, F.~Breyer, and S.~C.~J. Lee, ``{High-speed transmission over
  multimode optical fibers},'' in \emph{Proc.~Optical Fiber Communication
  Conference}, 2008, p. OWR2.

\bibitem{Hranilovic2003}
S.~Hranilovic and F.~R. Kschischang, ``{Optical intensity-modulated direct
  detection channels: Signal space and lattice codes},'' \emph{IEEE
  Transactions on Information Theory}, vol.~49, no.~6, pp. 1385--1399, 2003.

\bibitem{Karout2011GC}
J.~Karout, E.~Agrell, K.~Szczerba, and M.~Karlsson, ``Designing power-efficient
  modulation formats for noncoherent optical systems,'' in \emph{Proc.~IEEE
  Global Communications Conference}, 2011.

\bibitem{Karout2011IT}
------, ``Optimizing constellations for single-subcarrier intensity-modulated
  optical systems,'' \emph{\emph{submitted to} IEEE Transactions on Information
  Theory}, Jun. 2011, arXiv:1106.2819.

\bibitem{Hobook2005}
K.-P. Ho, \emph{{Phase-Modulated Optical Communication Systems}}.\hskip 1em
  plus 0.5em minus 0.4em\relax New York: Springer, 2005.

\bibitem{Walklin}
S.~Walklin and J.~Conradi, ``{Multilevel signaling for increasing the reach of
  10 Gb/s lightwave systems},'' \emph{Journal of Lightwave Technology},
  vol.~17, no.~11, pp. 2235--2248, 1999.

\bibitem{Simon1995}
M.~K. Simon, S.~M. Hinedi, and W.~C. Lindsey, \emph{{Digital Communication
  Techniques: Signal Design and Detection}}.\hskip 1em plus 0.5em minus
  0.4em\relax Englewood Cliffs, NJ: Prentice-Hall, 1995.

\bibitem{Cox2002}
C.~Cox and W.~S.~C. Chang, ``{Figures of merit and performance analysis of
  photonic microwave links},'' in \emph{{RF Photonic Technology in Optical
  Fiber Links}}, W.~S.~C. Chang, Ed.\hskip 1em plus 0.5em minus 0.4em\relax
  Cambridge University Press, 2002, ch.~1, pp. 1--33.

\bibitem{Agrawal2005}
G.~P. Agrawal, \emph{{Lightwave Technology: Telecommunication Systems}}.\hskip
  1em plus 0.5em minus 0.4em\relax New Jersey: John Wiley \& Sons, Inc., 2005.

\bibitem{Inan2009}
B.~Inan, S.~C.~J. Lee, S.~Randel, I.~Neokosmidis, A.~M.~J. Koonen, and J.~W.
  Walewski, ``Impact of {LED} nonlinearity on discrete multitone modulation,''
  \emph{Journal of Optical Communications and Networking}, vol.~1, no.~5, pp.
  439--451, Oct. 2009.

\bibitem{Wilson96}
S.~G. Wilson, \emph{{Digital Modulation and Coding}}.\hskip 1em plus 0.5em
  minus 0.4em\relax New Jersey: Prentice-Hall, 1996.

\end{thebibliography}
\end{document}